# Nonlinear Behavior of Josephson Traveling Wave Parametric Amplifiers

Claudio Guarcello, Felix Ahrens, Guerino Avallone, Carlo Barone, Matteo Borghesi, Luca Callegaro, Giovanni Carapella, Anna Paola Caricato, Iacopo Carusotto, Alessandro Cian, Alessandro D'Elia, Daniele Di Gioacchino, Emanuele Enrico, Paolo Falferi, Luca Fasolo, Marco Faverzani, Elena Ferri, Giovanni Filatrella, Claudio Gatti, Andrea Giachero, Damiano Giubertoni, Veronica Granata, Angelo Leo, Danilo Labranca, Carlo Ligi, Giovanni Maccarrone, Federica Mantegazzini, Benno Margesin, Giuseppe Maruccio, Renato Mezzena, Anna Grazia Monteduro, Roberto Moretti, Angelo Nucciotti, Luca Oberto, Luca Origo, Alex Stephane Piedjou Komnang, Sergio Pagano, Luca Piersanti, Alessio Rettaroli, Silvia Rizzato, Simone Tocci, Andrea Vinante, and Mario Zannoni

*Abstract*— Recent advancements in quantum technologies and advanced detection experiments have underscored the pressing need for the detection of exceedingly weak signals within the microwave frequency spectrum. Addressing this challenge, the Josephson Traveling Wave Parametric Amplifier (JTWPA) has been proposed as a cryogenic front-end amplifier capable of approaching the quantum noise limit while providing a relevant bandwidth. This research is centered on a comprehensive numerical investigation of the JTWPA, without resorting to simplifications regarding the nonlinearity of the essential components. Specifically, this study focuses on a thorough examination of the system, characterized by coupled nonlinear differential equations representing all components of the device. Proper input and output signals at the device's boundaries are considered. The analysis of the output signals undergoing the parametric amplification process involves a detailed exploration of phase-space dynamics and Fourier spectral analysis of the output voltage. This study is conducted while considering the parameters ruling the response of the device under pump and signal excitations. In addition to the expected signal amplification, the findings reveal that the nonlinear nature of the system can give rise to unforeseen phenomena, depending on the system's operational conditions, which include: the generation of pump tone harmonics, modulation of the signal gain, and incommensurate frequency generation—effects that are not easily accommodated by simplistic linearized approaches.

## I. INTRODUCTION

NOWADAYS, there has been considerable interest in superconducting quantum limited parametric amplifiers due to their relevance in various practical applications, e.g., quantum optics [1] [2], quantum measurement [3] qubit readout [4], and microwave single photon detection [5] [6] [7] [8] [9] [10]. Many of these applications involve the simultaneous multiple reading and the utilization of multiplexing strategies, needing large bandwidth amplifiers with minimal noise.

Quantum limited parametric amplifiers offer a clear advantage by providing substantial amplification while enabling the possibility of approaching and, in some cases, surpassing the

Manuscript receipt and acceptance dates will be inserted here. This work was supported by DARTWARS, a project funded by the Italian Institute of Nuclear Physics (INFN) within the Technological and Interdisciplinary Research Commission (CSN5) and by European Union's H2020-MSCA Grant Agreement No. 101027746, and by University of Salerno - Italy under the projects FRB19PAGAN, FRB20BARON, and FRB22PAGAN. *(Corresponding author: Claudio Guarcello)*

C. Guarcello, G. Avallone, C. Barone, G. Carapella, V. Granata and S. Pagano are with Physics Dept., University of Salerno, with INFN - Gruppo Collegato Salerno, and with CNR-SPIN Salerno section, 84084 Fisciano, Salerno, Italy (e-mail: cguarcello@unisa.it; guavallone@unisa.it; cbarone@unisa.it; gcarapella@unisa.it; vgranata@unisa.it; spagano@unisa.it).

F. Ahrens, A. Cian, D. Giubertoni, F. Mantegazzini, and B. Margesin are with Fondazione Bruno Kessler and with INFN - TIFPA, Via Sommarive, I-38123, Povo, Trento, Italy (e-mail: fahrens@fbk.eu; acian@fbk.eu; fmantegazzini@fbk.eu; margesin@fbk.eu).

M. Borghesi, M. Faverzani, A. Giachero, D. Labranca, R. Moretti, A. Nucciotti, L. Origo and M. Zannoni are with University of Milano Bicocca, Department of Physics, with INFN - Milano Bicocca, and with Bicocca Quantum Technologies (BiQuTe) Centre, Piazza della Scienza, I-20126 Milano, Italy. (e-mail: matteo.borghesi@unimib.it; marco.faverzani@unimib.it; andrea.giachero@unimib.it; d.labranca@campus.unimib.it; r.moretti9@campus.unimib.it; angelo.nucciotti@unimib.it; l.origo2@campus.unimib.it; mario.zannoni@unimib.it).

E. Ferri is with INFN - Milano Bicocca, Piazza della Scienza, I-20126 Milano, Italy. (e-mail: elena.ferri@unimib.it)

E. Enrico and L. Oberto are with INRiM - Istituto Nazionale di Ricerca Metrologica, Strada delle Cacce, I-10135 Turin, Italy and with INFN - TIFPA, Via Sommarive, I-38123, Povo, Trento, Italy (e-mail: e.enrico@inrim.it; l.oberto@inrim.it).

L. Fasolo and L. Callegaro are with INRiM - Istituto Nazionale di Ricerca Metrologica, Strada delle Cacce, I-10135 Turin, Italy (e-mail: l.fasolo@inrim.it; l.callegaro@inrim.it).

A.P. Caricato, A. Leo, G. Maruccio, A.G. Monteduro and S. Rizzato are with Mathematics and Physics Dept., University of Salento, and with INFN – Sezione di Lecce, 73100 Lecce, Italy (e-mail: annapaola.caricato@unisalento.it; angelo.leo@unisalento.it; giuseppe.maruccio@unisalento.it; annagrazia.monteduro@unisalento.it; silvia.rizzato@unisalento.it).

I. Carusotto is with CNR INO, and with Physics Dept., University of Trento, 38123, Povo (TN), Italy (email: iacopo.carusotto@ino.cnr.it).

A. D'Elia, D. Di Gioacchino, C. Gatti, C. Ligi, G. Maccarrone A.S. Piedjou Komnang, L. Piersanti, A Rettaroli, and S. Tocci are with INFN - Laboratori Nazionali di Frascati, Via Enrico Fermi, 00044 Frascati (RM), Italy (e-mail: alessandro.delia@lnf.infn.it; daniele.digioacchino@lnf.infn.it; claudio.gatti@lnf.infn.it; carlo.ligi@lnf.infn.it; giovanni.maccarrone@lnf.infn.it; luca.piersanti@lnf.infn.it; apiedjou@lnf.infn.it; Alessio.Rettaroli@lnf.infn.it; Simone.Tocci@lnf.infn.it).

P. Falferi and A. Vinante are with Fondazione Bruno Kessler, with CNR IFN, and with INFN - TIFPA, Via Sommarive, I-38123, Povo, Trento, (email: paolo.falferi@unitn.it; andrea.vinante@ifn.cnr.it).

R. Mezzena is with Physics Dept., University of Trento, and with INFN – TIFPA, I-38123, Povo, Trento, Italy (e-mail: renato.mezzena@unitn.it).

G. Filatrella is with Science and Technology Dept., University of Sannio, 82100 Benevento, Italy and with INFN - Gruppo Collegato Salerno, 84084 Fisciano (SA), Italy (e-mail: filatr@unisannio.it).

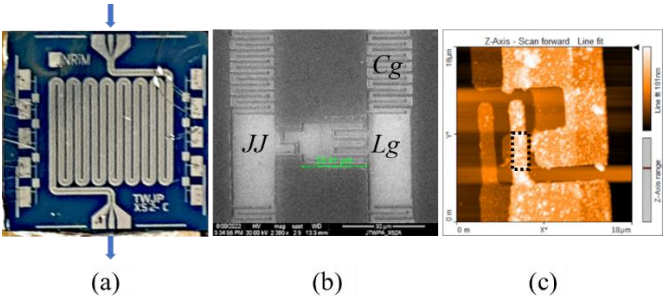

**Fig. 1.** The realized JTWPA: (a) the whole chip; (b) a SEM photograph of a single cell; (c) an AFM picture of one of the 990 JJs forming the JTWPA

quantum noise limit under specific conditions [11].

The fundamental principle of parametric amplification hinges on the generation of frequency mixing induced by the system nonlinearity. Practically, it entails the amplification of a weak signal through its interaction with a more robust pump tone, leading to a pump-to-signal energy transfer. The implementation of these devices frequently needs Josephson junctions (JJs), which, due to their nonlinear inductance, act as tunable nonlinear resonators, thus allowing adjustment of the resonance frequency to achieve optimized amplification.

## II. JTWPA

Josephson parametric amplifiers (JPAs) achieve power transfer from a robust pump tone to a weak signal through the inherent non-linearity of the JJ and a subsequent frequency mixing process. The interest reader can find a comprehensive review of recent JPAs developments in [12]. It's worth noting that, while JPAs can approach the quantum noise limit, they exhibit limitations in terms of bandwidth, dynamic range, and linearity [12] [13]. In particular, the limited bandwidth of JPAs restricts their feasibility. For instance, in applications like qubit readout, the capacity to read a few qubits per amplifier is a bottleneck in scaling up quantum processors [14] [15]. Additionally, in axion detection experiments, scaling the setup to include a series of cavities is not feasible [16] [17] [18].

In contrast, superconducting Traveling Wave Parametric Amplifiers (TWPAs) maintain a low quantum noise level while offering a broad bandwidth. Although the theoretical framework for these amplifiers is relatively aged [19] [20], they are currently experiencing a resurgence in popularity, with frequent new proposals and implementations emerging. Superconducting TWPAs operate by parametrically amplifying microwaves propagating along a transmission line containing nonlinear components, often realized using JJs. The current-dependent Josephson inductance plays a pivotal role in the parametric mixing and amplification process. By injecting a high-amplitude pump tone at frequency $f_{pump}$ and a weak signal to be amplified at frequency $f_{sign}$, the large-amplitude pump tone modulates the nonlinear Josephson inductance, facilitating the coupling between the pump and signal. This process may result in signal amplification and the generation of an idle tone, at frequency $f_{idle}$. Depending on the current dependence of the nonlinearity, two distinct operating regimes emerge: a four-wave mixing (4WM) mode, where the following relation holds: $2f_{pump} = f_{sign} + f_{idle}$, and a three-wave mixing (3WM) mode, where the relation between the involved frequencies is: $f_{pump} = f_{sign} + f_{idle}$, reflecting the conservation of energy in the photon mixing process. The latter mode arises when the nonlinear inductance term exhibits odd dependence on current and is typically obtained by biasing the line with a DC current and/or magnetic field.

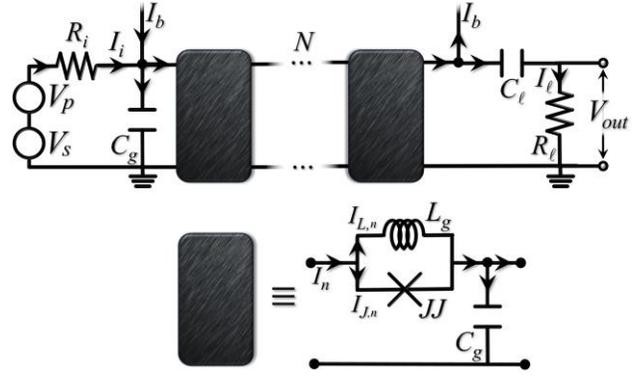

**Fig. 2.** Equivalent circuit of the analyzed JTWPA composed by $N$ identical replicas of the cell shown at the bottom.

This study focuses on the modeling of a specific JTWPA scheme, comprising a series of SQUIDs [21] [22] [23] [24] [25] [26] [27] and a capacitive shunt to ground. Typically, these devices offer nearly quantum-limited noise levels and a wide bandwidth (approximately 6 GHz centered at 7 GHz), albeit with limited gain (<20 dB) and dynamic range (<−90 dBm). For an in-depth description of JTWPAs, please refer to [28] [29].

In Fig. 1 is shown one of the realized JTWPAs, designed and fabricated at INRiM - Italy as part of the DARTWARS project [30] [31] [32] [33] [34]. Fig. 1a shows the whole chip, having lateral size of 1 cm, with the input port for the signal and pump indicated by the inward arrow and the output port by the outward arrow. The ports are connected, through suitable waveguides, to the rest of the amplification chain. The JTWPA is formed by 990 cells in series, each comprising a parallel connection between a *JJ* and an inductor, forming an RF-SQUID, and a grounding capacitor. These components are indicated by the *JJ*, $L_g$, and $C_g$ labels in Fig. 1b. The JJs were realized by e-beam evaporation of an Al/AlOx/Al trilayer using the shadow mask deposition technique. In Fig. 1c is shown an AFM image of one of the fabricated junctions. The junction area is outlined by the dotted rectangle. Typical obtained junction dimensions are 3.8 μm by 1.6 μm.

TABLE I
VALUES USED FOR THE PARAMETERS

| Name  | $N$ | $R_i$ | $R_l$ | $C_l$ | $C_g$ | $L_g$ | $R_J$ | $C_J$ | $I_J$ |
|-------|-----|-------|-------|-------|-------|-------|-------|-------|-------|
| units |     | Ω     | Ω     | nF    | fF    | pH    | kΩ    | fF    | μA    |
| value | 990 | 50    | 50    | 1     | 24    | 120   | 20    | 200   | 2     |

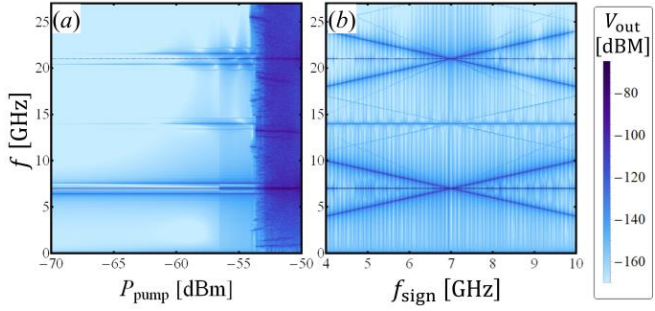

**Fig. 3.** Fourier spectra of the output voltage versus the pump power level for $f_{sign} = 6.42$ GHz (a) and versus the signal frequency for $P_{pump} = -55$ dBm (b). The color intensity scale represents the amplitude of the spectral components whose frequency can be read on the left vertical axis. The pump frequency is 7 GHz and the input signal power is -100 dBm.

## III. MODELLING

Figure 2 shows the electric circuit equivalent of the JTWPA analyzed in this study. The gray-shaded section within Fig. 2 shows to the generic $n^{th}$ cell within the JTWPA, while the top row delineates left and right boundary elements. The specific parameter values are derived from experimental data [31] and shown in Table 1. The arrows represent the injection points of an optional DC bias current $I_b$ used to modulate nonlinearities.

The JTWPA under consideration in this work comprises a series of $N$ identical cells, each housing an rf-SQUID. These cells consist of a JJ in parallel with a geometric inductance $L_g$ and a capacitor $C_g$ to ground. At the initial cell in the chain is applied the voltage of the pump and signal tones, denoted as:

$$V_i(t) = V_p(t) + V_s(t) =$$
$$= V_{pump} \sin(2\pi f_{pump} t) + V_{sign} \sin(2\pi f_{sign} t) \quad (1)$$

through an ideal linear transmission line, modeled by the input resistance $R_i$. The last chain element sees the load resistance $R_l$ through a capacitor $C_l$, inserted to block the DC current. The equations governing the behavior of a generic $n^{th}$ cell are:

$$I_n - I_{n+1} = \frac{dq_n}{dt} \quad (2)$$

$$I_n = I_{J,n} + I_{L,n} = I_{J,n} + \frac{\hbar}{2e} \frac{\varphi_n}{L_{g,n}} \quad (3)$$

$$\frac{\hbar}{2e} \frac{d\varphi_n}{dt} + \frac{q_n}{C_{g,n}} = \frac{q_{n-1}}{C_{g,n-1}}, \quad (4)$$

while the boundary conditions can be written are

$$\frac{dI_i}{dt} = \frac{1}{R_i} \frac{dV_i}{dt} - \omega_i(I_i + I_b - I_1) \quad (3)$$

$$\frac{dI_l}{dt} = -\omega_l \left[ \left(1 + \frac{C_{g,N}}{C_l}\right) I_l + I_b - I_N \right]. \quad (4)$$

Here: $\varphi$ is the Josephson phase difference; $\hbar$ and $e$ are the reduced Planck constant and electron charge; $q_n$ is the charge on the capacitor $C_{g,n}$; $I_n$, $I_{L,n}$ and $I_{J,n}$ are the total, the $L_{g,n}$ and the JJ currents respectively; $\omega_i = 1/(R_i C_g)$ and $\omega_l = 1/(R_l C_{g,N})$. The current $I_{J,n}$ is accounted for by the Resistively and Capacitively Shunted Junction (RCSJ) model [36]:

$$I_{J,n} = C_J \frac{\hbar}{2e} \frac{d^2 \varphi_n}{dt^2} + \frac{1}{R_J} \frac{\hbar}{2e} \frac{d\varphi_n}{dt} + I_c \sin \varphi_n, \quad (5)$$

where $I_c$ the Josephson critical current, and $C_J$ and $R_J$ are the junction capacitance and resistance, respectively.

These values yield a plasma frequency $f_p \approx 27.7$ GHz. The value of $R_J$ is estimated and chosen large enough to not affect the overall dynamics. Using the 2$^{nd}$ Josephson relation, $V = \frac{\hbar}{2e} \frac{d\varphi}{dt}$, the nonlinear inductance of a JJ, $L_J(I)$, for $I < I_c$, is:

$$L_J(I) = \frac{\hbar}{2e\, I_c} \frac{1}{\sqrt{(1-(I/I_c)^2)}} \quad (6)$$

In essence, the bias current serves as a control parameter for tuning the Josephson inductance. As $I$ increases, the inductance grows, eventually diverging for $I \to I_c$.

Theoretical approaches [36] [37] typically linearize $L_J(I)$ and consider simplified scenarios in lower-order approximations. In contrast, our approach addresses the problem in its entirety by solving a system of $N$ coupled differential equations [26], with appropriate boundary conditions. The system's dynamics are obtained by integrating these $N$ coupled differential equations using a finite-difference implicit method based on a tridiagonal algorithm, often used for numerical studying Josephson transmission lines [38] [39] [40]. The output signal is analyzed in its spectral components using an algorithm based on Mathematica® software. The chosen values for the normalized time step and integration time are $\Delta t = 0.01$ and $t_{max} = 20000$, respectively, corresponding to $\Delta t = 0.06\, ps$ and $t_{max} = 120\, ns$. This allows to integrate for a sufficient time to rule out transient phenomena while preserving stability of the integrator. The chosen methodology enables to analyze the complete signals in the time and frequency domains, the effect of each system parameter on the device performances, the effect of spatial changes of the parameters due to the imperfections that affect practical realization of devices, and the effects induced by intrinsic thermal noise. In this work, identical cells for the JTWPA are assumed. The effects of parameter value distributions [41] and of thermal noise will be explored in future work. Next sections present numerical integration results involving the JTWPA operating in 4WM mode, and analyze the effects of pump power, signal frequency, and of the variation of the transmission line parameters on the overall dynamics and, more specifically, on the amplifier gain.

## IV. RESULTS

Figure 3 shows two examples of Fourier spectra of output voltage by fixing the pump frequency to 7 GHz and the input signal power to -100 dBm. Figure 3(a) shows the output spectral power densities (in intensity scale) resulting by sweeping the pump power from -70 to -50 dBm and with $f_{sign} = 6.42$ GHz. The pump, signal, and idle peaks, and their harmonics, are clearly visible for all pump powers. Three regions are evident in the spectra, as also discussed in Ref. [42]:

*i)* $P_{pump} < -57$ dBm: a moderate gain ($< 5$ dB) is obtained.
*ii)* $P_{pump} \in [-57, -54]$ dBm: the signal gain increases up to ~10 dB, as well as the number of visible harmonics.
*iii)* $P_{pump} > -54$ dBm: the apparent gain is large, but the system enters a turbulent regime with a large noise rise at all $f_{sign}$.
Last situation, caused by the overly intense pump power, is certainly not suitable for an amplifier. Consequently, there is a limit to the maximum power available for amplification. It is worth nothing that this effect can be only seen from the numerical simulation of the full nonlinear system dynamics.

Figure 3(b) displays the output spectra obtained by setting $P_{pump}$=-55 dBm, corresponding to the average "clean" gain region, and sweeping $f_{sign}$ from 4 GHz to 10 GHz. The peaks corresponding to the pump, signal, and idle are clearly visible in all frequency range, with their first and second harmonics, denoting the large bandwidth of the amplifier. Also, some other lines are visible, corresponding to different combinations between the involved frequencies. These signals subtract power from the signal to be amplified and can generate instability. The vertical bands visible in Fig. 3(b) are due to gain ripples, which is discussed in more details in the following.

Figure 4 (a) shows the gain *vs* frequency response of the amplifier, when operated in the "clean" gain region, that is with $f_{pump} = 7$ GHz, $P_{pump} = -55$ dBm and $P_{sign} = -100$ dBm. All the other system parameters are as reported in Table I. The so-called "ripples" in the gain profile are clearly visible, see the inset. They are due to variable impedance along the transmission line that do not well matches the input and output impedances and generate interference between forward and backward propagating waves undergoing multiple reflections [41] [43]. These ripples have been described as Fabry-Pérot-like resonances [26] [44] with a bandwidth inversely proportional to the waveguide length. Recently, an approach based on the theory of coupled modes [36], successfully predicts the formation of these ripples by considering the interactions of reflected waves [45]. In the investigated scheme, the emerging ripple pattern depends on parameter values and profoundly affects both the gain and bandwidth of the device.

To better clarify this effect, in Figs. 4 (b)-(d) the gain response has been computed by varying the system parameters. Figures 4 (b) and (e) are obtained by setting $C_g$ to 16 pF and 32 pF, respectively. A larger capacitance to ground increases the average gain, but also increases the number of ripples and reduces the bandwidth. Moreover, a further increase of the capacitance triggers a chaotic system response (data not shown). Figure 4 (d) illustrates what happens when decreasing $R_J$ from 20 kΩ to 1 kΩ, i.e., increasing the dissipation being the damping parameter equal to $\alpha=1/(R_J C_J f_p)$. A significant reduction of the gain is evident, although it is fair to point out that at the usual working temperatures of JTWPAs, i.e. a few tens of mK, such small resistances are unrealistic. Figures (e) and (f) shows effects of changing the loop inductance $L_g$ to 100 pH and 130 pH, respectively. In the former case a suppression of ripples is obtained, together with a reduction in average gain, while in the latter case higher gains are achieved, but at the cost of instability in the higher-gain regions.

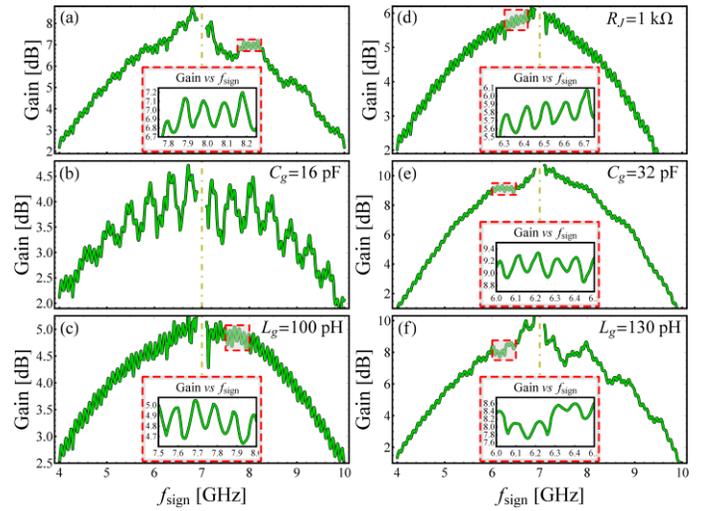

**Fig. 4.** Frequency response of the JTWPA and effect of change of parameters. The signal power is fixed to -100 dBm, the pump frequency to 7 GHz, and power to -55 dBm. The insets detail the gain ripples. The parameters in (a) are as in Table I, while in other panels only one parameter value changes from this table, i.e., (b) $C_g = 16$ pF, (c) $C_g = 32$ pF, (d) $R_J = 1$ kΩ, (e) $L_g = 100$ pH, and (f) $L_g = 130$ pH.

V. CONCLUSION

JTWPA are promising superconducting devices for wide-band ultra-weak signal amplification in the microwave spectral region. Semi analytic and simple numeric approaches, based on approximations in the nonlinearity and on limited frequency components, can reproduce most but not all the experimental features. Using a direct numerical simulation of the complete system, it has been shown that, beside the expected signal amplification and depending on the system operation, unwanted effects emerge (such as pump tone harmonics, incommensurate frequency generation, and ripple formation in the gain profile), which are not easily accounted for in simpler approaches. Moreover, the impact of the main system parameter on the response of the system has been investigated, making it possible to establish "threshold" values below which chaotic effects, detrimental to use as an amplifier, can be safely avoided.

Next steps will be investigation of 3WM and RPM configurations to closely reproduce experimental results and understand sources of nonidealities. The effects of intrinsic fluctuations in JJs, which are important in determining their switching [46][47][48], and their combined effect with the nonlinear dynamics will also be addressed as possible source of additional noise. Moreover, other schemes can be explored, for instance including two JJs in parallel in the rf-SQUID to achieve higher operation frequencies in "optical modes" [49].